\documentclass[twocolumn,amsmath,showkeys,prb]{revtex4-1}
\usepackage[bookmarks=false]{hyperref}
\usepackage{dcolumn}
\usepackage{bm}
\usepackage[T1]{fontenc}
\usepackage{amsmath}
\usepackage{graphicx}
\usepackage{tabularx}
\usepackage{multirow}
\usepackage[usenames, dvipsnames]{color}
\usepackage{hyperref}
\hypersetup{colorlinks=true, linkcolor=blue, citecolor=red, urlcolor=magenta, pdftitle={Topical Review}, pdfauthor={E. Bousquet, H. Hamdi}}

\begin{document}

\title{First-principles re-investigation of bulk WO$_3$}
\author{Hanen Hamdi$^{1}$, Ekhard K. H. Salje$^{2}$, Philippe Ghosez$^{1}$ and Eric Bousquet$^{1}$}
\affiliation{$^1$Theoretical Materials Physics, Q-MAT, CESAM, Universit\'e de Li\`ege, B-4000 Sart Tilman, Belgium}
\affiliation{$^2$ Department of Earth Sciences, University of Cambridge, Downing Street, Cambridge CB2 3EQ, United Kingdom}
\date{\today}

\begin{abstract}
Using first-principles calculations, we analyze the structural properties of tungsten trioxide WO$_3$. Our calculations rely on density functional theory and the use of the B1-WC hybrid functional, which provides very good agreement with experimental data.
The hypothetical high-symmetry cubic reference structure combines several ferroelectric and antiferrodistortive (antipolar cation motions, rotations and tilts of oxygen octahedra) structural instabilities. The instability related to antipolar W motions combines with those associated to oxygen rotations and tilts to produce the biggest energy reduction, yielding a $P2_1/c$ ground state. This non-polar $P2_1/c$ phase is only different from the experimentally reported $Pc$ ground state by the absence of a very tiny additional ferroelectric distortion. The calculations performed on a stoichiometric compound so suggest that the low temperature phase of WO$_3$ is not intrinsically ferroelectric and that the reported ferroelectric character might arise from extrinsic defects such as oxygen vacancies. Independently, we also identify never observed $R3m$ and $R3c$ ferroelectric phases with large polarizations and low energies close to the  $P2_1/c$ ground state, which makes WO$_3$ a potential antiferroelectric material.
The relative stability of various phases is discussed in terms of the couplings between different structural distortions, highlighting a very complex interplay involving improper-like couplings up to fourth order in the energy expansion in the cubic phase.

\end{abstract}
\maketitle

\section{Introduction}
Tungsten trioxide, WO$_3$, has been extensively studied due to its very attractive and rich properties for technological applications.
WO$_3$ and its derivatives H$_x$WO$_3$ and WO$_{3-x}$ are electrochromic,\cite{SALJE1978231, Niklasson2007,SatyenK2008, Baetens201087, Lee2006, Subrahmanyam2007266} thermocromic,\cite{Sella19981477, Zhao1997} and superconducting.\cite{Zhao1997, Deb2008245,Alison1998, AAird1998, Ekhard1976, Shengelaya1999, Barzykin2002,Kim2010}
It has been envisaged that WO$_3$  may become one of the best materials for electrochromic applications such as in energy-efficient windows, smart glasses,
antiglare automobile rear-view mirrors, sunroofs, displays, or even tunable photonic crystals\cite{GranqvistCG2000} and to reduce photocorrosion.\cite{Gillaspie2010} 
The wide variety of the underlying electronic instabilities for these properties is
mirrored by a multitude of related structural instabilities, which were investigated ever since 1975\cite{Salje1975, SaljeEK1975} and refined later.\cite{KLocherer1999, Salje1997, LochererKR1999, Vogt1999, Christopher2002, MBoulova2002} 

All known WO$_3$ phases are characterized by very large distortions of the archetypal perovskite structure so that even the notion of octahedra tilts is to be taken with some caution.
The WO$_6$ octahedra are so largely distorted that the variance of W--O distances in any observed structure is far greater than in most other known perovskite structures.\cite{Diehl1978, Woodward1995, Woodward1979, Sigetosi1960, Loopstra1966}
In this paper we make the fundamental connection between the electronic and structural structure properties of WO$_3$ and show that most, if not all, structural instabilities can be derived from a careful analysis of its intrinsic cubic phonon instabilities, despite these large deformation amplitudes.

The structural properties can be summarized as follows: WO$_3$ shows no proper melt, surface melting of crystalline material occurs at 1746 K.
Crystal growth proceeds typically by sublimation and gas transport at temperatures below 1400K.
At the highest temperatures the structures is tetragonal (space group  $P4/nmm$) with strong antiferrodistortive (AFD) cation movements so that the WO$_6$ octahedra are strongly distorted\cite{Kehl1952} in an anti-polar pattern.
Additional rotational octahedral distortions condense in addition to the initial tetragonal displacements when lowering the temperature.
They further reduce the symmetry from tetragonal to orthorhombic, monoclinic, triclinic, and finally to a second monoclinic phase.
A structural sequence, which contains phases stabilized by temperature, is given in Fig.1.
A monoclinic phase ($P2_1/n$)\cite {LochererKR1999} and a triclinic phase  $P\bar{1}$  exist at room temperature.\cite{Woodward1995, Diehl1978}
At higher temperatures, Vogt\cite{Vogt1999} et al and Locherer \textit{et al.}\cite{KLocherer1999} concluded a transition from $Pbcn$ to the $P4/ncc$ phase and Howard \cite{Howard2002} observed an intermediate $P2_1/c$ phase.
Locherer and Woodward found an additional transition from $P4/ncc$ to $P4/nmm$ near at 980 K to 1200 K.
Below room temperature, Salje et al\cite{Salje1997} reported a transition from the triclinic  $P\bar{1}$  phase to a polar phase ($Pc$) with no further transitions down to 5K.

WO$_3$ occurs (almost) always as oxygen deficient WO$_{3-x}$ with a metal-insulator (MI) transition to a metallic phase for high concentrations of oxygen vacancies or doping with alkali metals.
Superconductivity occurs in the metallic phase\cite{aird1998a} even if the reduced regions are restricted to nano-scale twin boundaries.
Bulk superconductivity in WO$_{3-x}$ was found in a tetragonal phase with space group  $P\bar{4}2_1m$.\cite{aird1998} 
(Bi-) polaronic electron transport is a characteristic property of WO$_{3-x}$.\cite{SCHIRMER1980333, Schirmer1980, SALJE1979237, Salje1984}
\begin{figure}
\centering
\includegraphics[width=8cm,keepaspectratio=true]{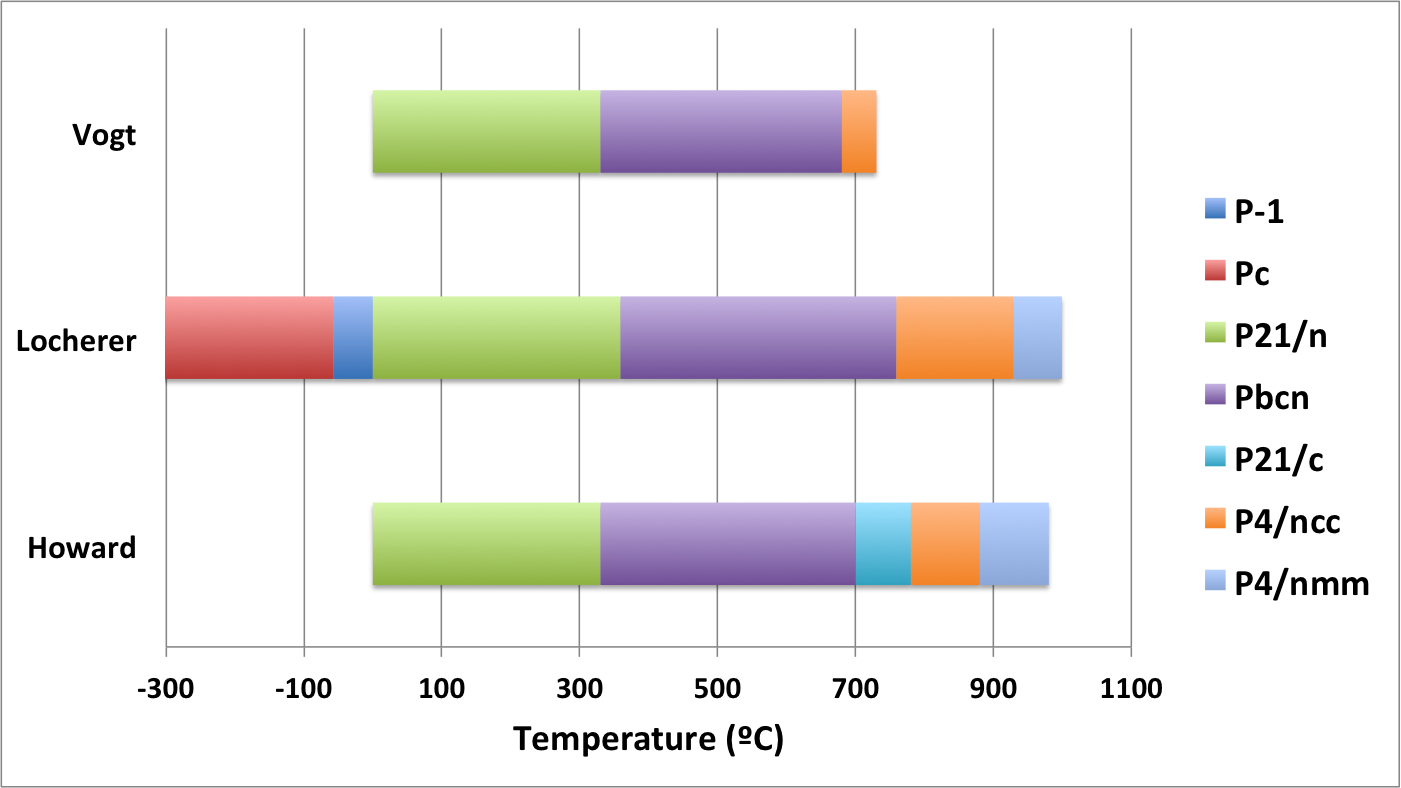}
\caption{(Color online) Schematic summary of the temperature phase diagram WO$_3$ as reported by three main experimental sources (Vogt from Ref.~\onlinecite{Vogt1999}, Locherer from Ref.~\onlinecite{LochererKR1999} and Howard from Ref.~\onlinecite{Howard2002}).}
\label{fig:exp_diag}
\end{figure}

Numerous first-principles studies were performed on WO$_3$ in order to characterize its electronic structure (bulk, thin films and cluster phases\cite{GonzalezBorrero2010, Johansson2013, ping2013, Wijs1999, tsipis2000, stachiotti1997, cora1996}) the role of oxygen vacancies,\cite{WangFenggong2011, migas2010, wang2011, karazhanov2003}
and cation doping.\cite{walkingshaw2004, tosoni2014, HjelmAnders1996, ingham2005, huda2009, cora1997, huda2008, chen2013}
In this paper we focus on the effect of the electronic structure on the structural stabilities and metastabilities of the various phases.
We show that the hybrid functional B1-WC is preferable for the study of the electronic and structural properties of WO$_3$ over previous approaches.
In the first section we check the validity of the B1-WC functional against six known crystallographic phases.
The sole disagreement exists for the crystallographic $\varepsilon$ phase ($Pc$), which yields a higher symmetric $P2_1/c$ structure.
In the second section we analyze metastable phases starting from the phonon dispersions of the hypothetic cubic phase, identifying the main phonon instabilities.
We then condense various possible combinations of these unstable modes in order to reproduce all experimentally observed structures.
This procedure also reveals two new polar phases that are close in energy to the ground-state.

\section{Computational details}
As a preliminary test, we performed calculations on WO$_3$ using several approaches and codes.
We tested the norm-conserving pseudopotentials and projected augmented waves with ABINIT\cite{abinit2009} and VASP\cite{vasp1,vasp2} codes
working within local density approximation\cite{TeterMichael1993} (LDA) and different generalized gradient approximations (GGA) including PBE,\cite{PerdewJohnP.1996} PBEsol\cite{perdew2008} and Wu-Cohen\cite{WuZhigang2006} for the exchange-correlation functional.
We also tested the localized orbital approach and the B1-WC\cite{BilcDI.2008} functional as implemented in the CRYSTAL code.\cite{CausaM1987}

As reported by Wang \textit{et al.},\cite{wang2011} we found that PBEsol and Wu-Cohen GGA functionals give good agreement for the structural properties but underestimate the electronic band gap, while the hybrid functionals give good agreement for both the structure and the band gap.
Wang \textit{et al.}\cite{wang2011} investigated B3LYP, PBE0 and HSE06 hybrid functionals and found that the HSE06 gives the best agreement with experiments.
In our study, the B1-WC  hybrid functional also gives good results, often better compared with the HSE06 functional.
We note that the HSE06 functional was used through the plane wave VASP code, which is highly computer time consuming while the B1-WC with the localized orbital scheme of CRYSTAL is more computationally efficient.

In what follows, we mainly focus on the results obtained with the B1-WC functional and the CRYSTAL code.
We have used the all-electron double-$\zeta$ basis sets for the oxygen atoms and small core Detlev Figgen pseudo-potentials,\cite{Figgen2009} associated with double-$\zeta$ valence basis sets for tungsten.
We performed full structural relaxations with a convergence criteria on the root-mean-square of the gradient and displacements smaller than $5\times10^{-4}$ hartree/bohr and $5\times10^{-4}$ bohr respectively.
The electronic self-consistent calculations were converged until the difference of the total energy was smaller than 10$^{-9}$  Hartree.
The phonon frequencies and Born effective charges were computed using frozen phonon numerical differences\cite{PascaleF2004, ZicovichWilson2004} and the electric polarization through the Berry phase technique.\cite{kingsmith1993}
The integration in the Brillouin zone has been performed with a $8\times8\times8$ grid of k-points for the cubic unit cell and a  $4\times4\times4$ grid for cells doubled in the three directions with respect to the cubic one.

\section{Analysis of the experimental phases}
Several DFT studies of WO$_3$ have been performed previously\cite{tosoni2014,ping2013,WangFenggong2011,wang2011,migas2010,huda2009, huda2008,ingham2005,chatten2005,walkingshaw2004,Wijs1999,cora1997,cora1996,HjelmAnders1996}
essentially focusing on the main and most common phases and on the electronic structure analysis with and without oxygen vacancies.
A detailed analysis of the complex structural phase diagram of WO$_3$ is thus missing while a microscopic knowledge of the origin of these different phases would be extremely valuable to understand the unique properties of WO$_3$.

In this section we start by characterizing the different phases of WO$_3$ observed experimentally to validate our approach and we will discuss the possible origin of the $Pc$ phase.
We will also analyze the electronic structure of these phases and we will discuss how the B1-WC compares with the previous studies.

\subsection{Structural and crystallographic analysis}
In Table \ref{tab:positions} we compare our calculated crystallographic data of the  $P4/nmm$, $P4/ncc$, $Pbcn$, $P2_1/n$, $P\bar{1}$  and $P2_1/c$ phases against the experimental measurements.
Because the $P2_1/c$ phase is not observed experimentally at low temperature, we compare it with the closely related experimental $Pc$ phase.

Our calculations of the $P4/nmm$ phase are in very good agreement with the observed cell parameters and the atomic positions.
The $P4/nmm$ phase consists of highly distorted WO$_3$ octahedra where the W--O bonds dimerize in opposite direction along the [110] perovskite direction.
This W--O dimerization forms local dipole-moments that are aligned along the [001] direction and anti-aligned along the [110] direction, so that the total dipole-moment cancels.
The crystallographic unit cell is elongated along the [001] direction and compressed along the [100] and [010] directions.

The $P4/ncc$ phase shows additional octahedra rotations around the $z$  axis (a$^0$a$^0$c$^-$ in the Glazer notation\cite{glazer1972}), which induces a cell doubling along the [001] direction.
The calculated $c$ cell parameter and the $z$ component of the atomic positions are in good agreement with experiments while the in-plane displacements are less well reproduced (Tab.~\ref{table1}).
The calculated $a$ and $b$ cell parameters are smaller than in experiments by 0.1 \AA\ and the deviation from the tetragonal O$_2$ position are about two times larger than observed.
We clearly overestimate the  a$^0$a$^0$c$^-$  distortions.
Note, however, that our calculations ignore thermal effects while experiments were performed at high temperatures.\cite{cochran71}
A possible comparison would be to extrapolate the experimental data for limited high temperature intervals to 0K but such data are not available for the $P4/ncc$ phase.

The orthorhombic $Pbcn$ phase can be characterized by its additional octahedra rotation about the crystallographic $y$ axis ($a^0$b$^+$a$^0$).
We find a similar overestimate of the octahedra distortions as for the $P4/ncc$ phase while the calculated cell parameters are underestimated with respect to experiments. We note that the anti-polar distortions along the z axis compares well with experiments for $P4/nmm$, $P4/ncc$ and $Pbcn$.

The $P2_1/n$ structure contains an additional octahedra rotation around the crystallographic $x$ axis (a$^-$b$^+$c$^-$).
The calculated cell volume is slightly too small ($+$0.7\%, $-$0.7\% and $-$1.9\% for $a$, $b$ and $c$ cell parameters respectively)  and the oxygen motions related to the octahedra tilt are overestimated (Table~\ref{table1}).

The $P\bar{1}$  phase is similar to the $P2_1/n$ phase if one replaces the in-phase rotation by an out-of-phase rotation with the pattern a$^-$b$^-$c$^-$.
The distortions are anisotropic in all three directions, which causes the cell to be triclinic with the angles $\alpha$, $\beta$ and $\gamma$ close to 90$^\circ$.
The calculated $a$, $b$ and $c$ cell parameters deviate from experiments by  $+$0.4\%, $-$1.1\% and $-$0.8\% respectively.

The largest apparent deviation is in the $\varepsilon$  ($Pc$) phase.
The $Pc$ phase is related to the $P2_1/c$ phase through an additional polar distortions along the c axis.
While relaxing the low temperature $Pc$ phase we observed that the system always relaxes back in to the higher symmetric $P2_1/c$ structure.
Wijs \textit{et al} using LDA and GGA exchange correlation functionals\cite{Wijs1999} found a similar effect.
In Table~\ref{table1} we compare our calculated atomic positions and cell parameters of the $P2_1/c$ phase with the experimentally determined $Pc$ phase.
The deviations are surprisingly small ($+$0.2\%, -0.2\% and -0.5\% for $a$, $b$ and $c$ lattice parameters) and even smaller differences for the atomic positions.
To further assess the dynamical stability of the $P2_1/c$ phase with respect to a potential $Pc$ ground state, we computed the zone-center phonons and did no observed any unstable mode.
The lowest polar mode has a frequency of 158 cm$^{-1}$ and is far from being unstable.
We also checked whether a soft polar mode can be generated by increasing the cell volume but did not observe any polar instability.
We thus follow the argument by Wijs \textit{et al.}\cite{Wijs1999}  that polarity in the experimental study may be stabilized by the presence of oxygen vacancies or by another extrinsic parameter.
Comparing the structural parameters obtained with other hybrid functionals PBE0, B3LYP and HSE06  reported by Wang et al.\cite{WangFenggong2011}, we find close agreement with a smaller error margin for B1-WC.
B1-WC gives a much better agreement for the $P2_1/c$ phase with experimental data than using the three hybrid functionals tested by Wang et al.:  HSE06, B3LYP and PBE0 with errors of $+$0.6\%, $+$1.3\% and $+$0.2\% on $a$, $+$2.1\%, $+$2.5\% and $+$0.6\% on $b$ and $+$0.1\%, $+$3.0\%, $+$1.7\% on the $c$ parameter.
We notice, however, that the B1-WC often underestimates cell parameters while the three other hybrid functionals overestimate the cell parameters of WO$_3$.

\begin{table*}[htbp!]
\caption{Calculated lattice parameters and Wyckoff positions of distorted WO$_3$
phases fully relaxed  with the  B1-WC functional. For each phase, we specify the space group and
the experimental parameters are reported for comparison}
\label{tab:positions}

\begin{tabular}{cc}

\begin{tabular}{ c c c c c c c}
\hline
\hline 
 $P4/nmm$ & \multicolumn{3}{c}{Present} & \multicolumn{3}{c}{ Exp.\cite{Howard2002}}  \\ 
 \hline
  & a & b & c & a & b & c \\
  & 5.29 & 5.29 & 3.93 & 5.29 & 5.29 & 3.92 \\
  &   x  &  y  &  z  &    x  &  y  &  z   \\
   W$_1$ (2c)  &  0.25  & 0.25    &  -0.064     &  0.25   & 0.25    & -0.066  \\      
   O$_1$ (2c) &  0.25  & 0.25    &  0.49       &  0.25   & 0.25    &  0.49   \\
   O$_2$ (4d)  &  0   & 0     &  0        &  0    & 0     &  0   \\
\hline
\hline
$$P4/ncc$$ & \multicolumn{3}{c}{Present} & \multicolumn{3}{c}{ Exp.\cite{Howard2002}}  \\ 
 \hline
   & a & b & c & a & b & c \\
  & 5.17 & 5.17 & 7.86 & 5.27 & 5.27 & 7.84 \\
  &   x  &  y  &  z  &    x  &  y  &  z   \\
   W$_1$ (4c)  &  0.25  & 0.25    &  0.2849     &  0.25   & 0.25    & 0.2832  \\      
   O$_1$ (4c)  &  0.25  & 0.25    &  0.0057     &  0.25   & 0.25    & 0.003   \\
   O$_2$ (8f) &  0.057 & -0.057  &  0.25       &  0.025  & -0.025  & 0.25     \\
\hline
\hline
$Pbcn$ & \multicolumn{3}{c}{Present} & \multicolumn{3}{c}{ Exp.\cite{Vogt1999}}  \\ 
\hline
   & a & b & c & a & b & c \\
   & 7.28 & 7.52 & 7.68 & 7.33 & 7.57 & 7.74 \\
   &   x  &  y  &  z  &    x &  y  &  z   \\
   W$_1$ (3d)   &  0.251  & 0.026   &  0.28     &  0.252   & 0.029  & 0.283  \\      
   O$_1$ (3d)   &  -0.001 & 0.043   &  0.215   &  -0.002  & 0.032  & 0.221   \\
   O$_2$ (3d)   &  0.293   & 0.259   &  0.259    &  0.283   & 0.269  & 0.259   \\
   O$_3$ (3d)   &  0.287   & 0.010  &  0.006   &  0.280   & 0.013  &  0.002   \\
   \hline
\end{tabular}

&

\begin{tabular}{ c c c c c c c}
\hline
\hline
  $P\bar{1}$  & \multicolumn{3}{c}{Present} & \multicolumn{3}{c}{ Exp.\cite{Diehl1978}}  \\ 
\hline
  & a & b & c & a & b & c \\
  & 7.33 & 7.44 & 7.61 & 7.30 & 7.52 & 7.67 \\
  &$\alpha$ & $\beta$ & $\gamma$ & $\alpha$ & $\beta$ & $\gamma$ \\
  & 88.64 & 91.02 & 91.01 & 88.81 & 90.92 & 90.93 \\
  &   x     &  y       &  z         &    x  &  y &  z   \\          
   W$_1$  (2i)    &   0.2603 & 0.0172 & 0.2826 &  0.2566 & 0.0259   &  0.2850     \\    
   W$_2$  (2i)   &   0.2540 & 0.5210 & 0.2183 &  0.2502 & 0.5280   &  0.2158    \\
   W$_3$  (2i) &   0.2397 & 0.0228 & 0.7793 &  0.2438 &  0.0313  & 0.7817    \\
   W$_4$  (2i)  &   0.2456 & 0.5268 & 0.7216 &  0.2499 & 0.5338   &  0.7190   \\
   O$_1$  (2i)  &   0.0015 & 0.0395 & 0.2074 &  0.0007 & 0.0386   &  0.2100    \\
   O$_2$  (2i)  &   0.5022 & 0.5406 & 0.2115 &  0.5038 & 0.5361   &  0.2181   \\
   O$_3$  (2i)  &   0.0026 & 0.4582 & 0.2897 &  0.0076 & 0.4660   &  0.2884    \\ 
   O$_4$  (2i)  &   0.5012 & -0.0398 & 0.2906 &  0.4972 & -0.0362  & 0.2878   \\
   O$_5$  (2i)  &   0.2892 & 0.2571 & 0.2836 &  0.2851 & 0.2574 & 0.2870       \\
   O$_6$  (2i)  &   0.2081 & 0.7575 & 0.2174 &  0.2204 & 0.7630 & 0.22232    \\ 
   O$_7$  (2i)  &   0.2098 & 0.2569 & 0.7232 &  0.2186 & 0.2627 & 0.7258     \\
   O$_8$  (2i)  &   0.2927 & 0.7575 & 0.7772 &  0.2840 & 0.7583 & 0.7679      \\ 
   O$_9$  (2i)  &   0.2911 & 0.0383 & 0.0060 &  0.2943 & 0.0422 & -0.0002      \\
   O$_10$ (2i)  &   0.2889 & 0.5389 & 0.4941 &  0.2971 & 0.5446 & 0.4982     \\
   O$_11$ (2i)  &   0.2108 & 0.4767 & -0.0061 &  0.2096 & 0.4820  & -0.0072    \\ 
   O$_12$ (2i)  &   0.2090 &-0.0242 & 0.5063  &  0.2088 & 0.9830  & 0.5051      \\
\hline
\end{tabular}

\\
 & \\

\begin{tabular}{ c c c c c c c}
 \hline
  \hline
  $P2_1/n$ & \multicolumn{3}{c}{Present} & \multicolumn{3}{c}{ Exp.\cite{Howard2002}}  \\ 
 \hline
  & a & b & c & a & b & c \\
  & 7.35 & 7.48 & 7.54 & 7.30 & 7.53 & 7.69 \\
  &$\alpha$ & $\beta$ & $\gamma$ & $\alpha$ & $\beta$ & $\gamma$ \\
  & 90 & 91.31 & 90 & 90 & 90.85 & 90 \\
  &   x     &  y       &  z         &    x  &  y  &  z   \\ 
   W$_1$  (4e)          &  0.2720  & 0.0074  &  0.2790   &  0.2528   & 0.02600  & 0.28550  \\      
   W$_2$  (4e)          &  0.2270  & 0.0133  &  0.7750   &  0.24970  &  0.034410  & 0.78050     \\
   O$_1$  (4e)          &  0.0043 &  0.0410  &  0.2165   &   0.0003  & 0.0337    & 0.2122 \\
   O$_2$  (4e)          &  -0.0056 & 0.4576   &  0.2170  &   -0.0011 & 0.4632  & 0.2177 \\
   O$_3$  (4e)          &  0.2883 & 0.2534   &  0.2924   &  0.28430  & 0.25980 & 0.2852 \\
   O$_4$  (4e)          &  0.2029 & 0.2530   &  0.7198   &  0.2080  & 0.2588   & 0.73320 \\
   O$_5$  (4e)          &  0.2795  &  0.0385  & 0.0059   & 0.28560 &  0.0410   & 0.0041       \\ 
   O$_6$  (4e)          &  0.2790   & 0.4630   & -0.0047 & 0.28410 & 0.48680   &-0.0056         \\ 
   \hline
\end{tabular}

&

\begin{tabular}{ c c c c c c c}
\hline
\hline
  $P2_1/c$ & \multicolumn{3}{c}{Present} & \multicolumn{3}{c}{ Exp. ($Pc$)\cite{SaljeEkhardKH1997}}  \\ 
\hline 
  & a & b & c & a & b & c \\
  & 5.26 & 5.15 & 7.62 & 5.28 & 5.16 & 7.66 \\
  &$\alpha$ & $\beta$ & $\gamma$ & $\alpha$ & $\beta$ & $\gamma$ \\
  & 90 & 91.78 & 90 & 90 & 91.75 & 90 \\
  &   x     &  y       &  z         &    x  &  y  &  z   \\
   W$_1$ (2a)          & -0.0093 & -0.0173  &  0.6843    &  -0.0099    &  -0.02   & 0.6743  \\      
   W$_2$  (2a)         &  0.5011 & 0.4827   &  0.7530    &  0.5        &   0.4710 & 0.75     \\
   O$_1$  (2a)         &  0.4975 & 0.5769   & -0.0245    & 0.4920      &   0.5780 & -0.0230 \\
   O$_2$ (2a)         &  0.2087 & 0.2891   &  0.1794    &  0.2130     &   0.2890 & 0.1830 \\
   O$_3$ (2a)         &  0.2830 & 0.7891   &  0.2580    &  0.2830     &   0.7860 & 0.2590 \\
   O$_4$  (2a)         &  0.6999 & 0.2090   &  0.1795    &  0.7050     &   0.2070 & 0.1820 \\
   O$_5$  (2a)         &  0.7918 & 0.7090   &  0.2579    &  0.7960     &   0.7110 & 0.2610 \\ 
   O$_6$  (2a)        & -0.0058 & 0.0769   &  0.4630    & -0.0058     &   0.073 & 0.4616 \\
   \hline
\end{tabular}

\end{tabular}
\label{table1}

\end{table*}

\subsection{Electronic structure}

In Table \ref{tab:gaps} we compare the calculated electronic structures for the hypothetical cubic,  $P4/nmm$, $P4/ncc$, $Pbcn$, $P2_1/n$, $P\bar{1}$  and $P2_1/c$ phases and compare them with the experiments and previous DFT calculations using PBE0, HSE06 and B3LYP hybrid functionals
and GW.
For the $P2_1/n$ and $P\bar{1}$  phases,  experimental data coincide with the B1-WC band gaps. The B1-WC results are similar to those obtained with the HSE06 functional while the PBE0 gives a slightly smaller gap energy and B3LYP larger values. The B1-WC band gap is closest to the results of GW calculations, an agreement also observed for the  $P2_1/c$ phase.

\begin{figure}[h]
 \centering
 \includegraphics[width=8.5 cm,keepaspectratio=true]{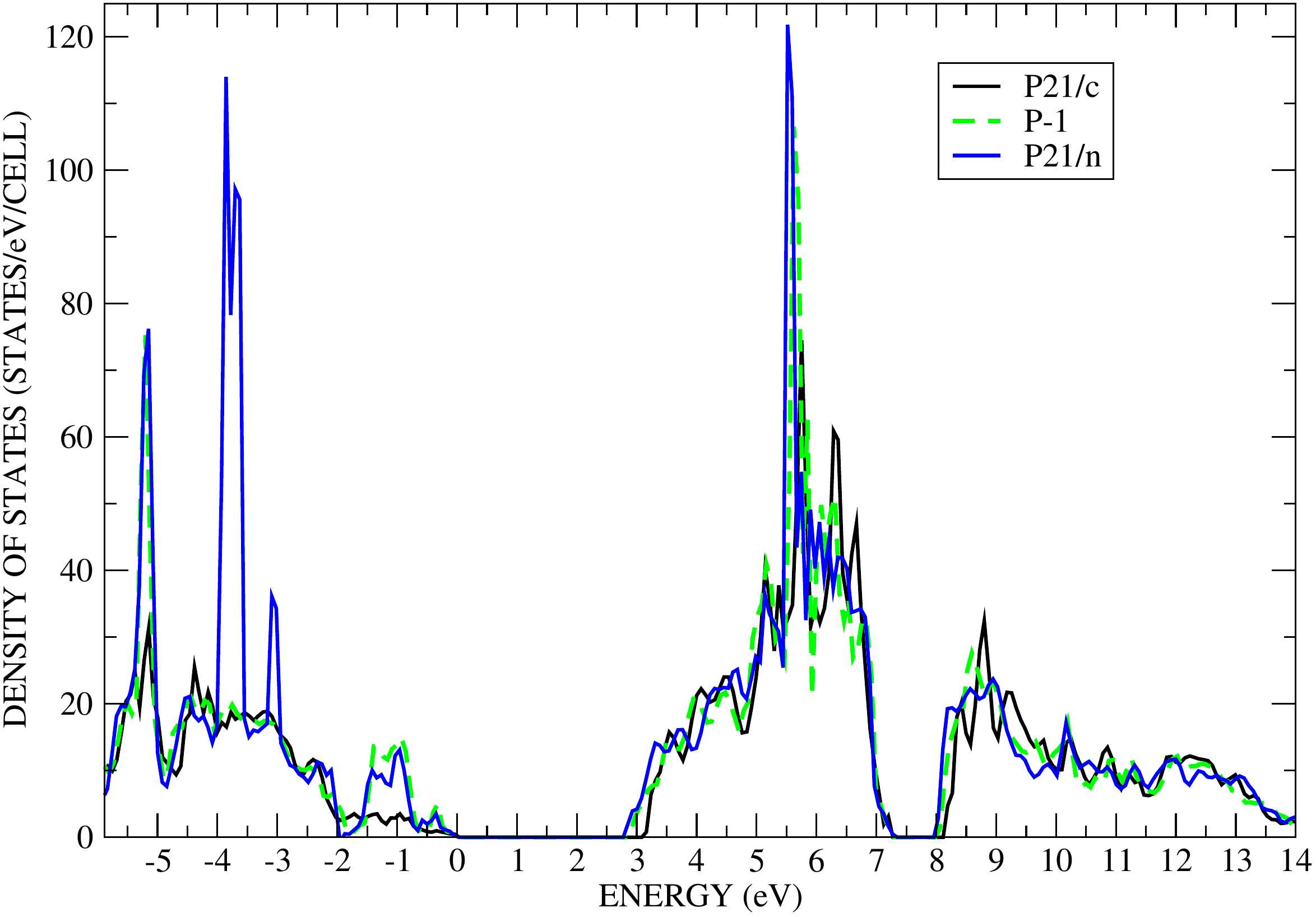}
 \caption{(Color online) Calculated density of states of the $P\bar{1}$ , $P2_1/n$ and $P2_1/c$ phases with the B1-WC functional.}
 \label{fig:DOS}
\end{figure}

Comparing the trend of band gaps between the different phases, we find that the band gap increases from the cubic (1.5 eV) to the lower symmetry phases (Table~\ref{tab:gaps}).
This means that both the anti-polar and octahedra tilt distortions increase the band gap of WO$_3$.
The calculated electronic gap energies is in reasonable agreement with the experimental values for the three
low-temperature structures: Eg = 2.85 eV for the room temperature monoclinic phase $P2_1/n$, Eg = 2.98 eV for the triclinic phase $P\bar{1}$  and Eg = 3.28 eV for the monoclinic phase $P2_1/c$.
Figure \ref{fig:DOS} shows the density of states (DOS) of these three phases to demonstrate their similarity.

\begin{table}[htbp!]
\caption{Electronic band gap (in eV) of different phases of WO$_3$ as calculated in the present work with the B1-WC hybrid functional. We compare with the results of previous hybrid functional calculations (PBE0, HSE06 and B3LYP)), the GW approach and experiments.}
\label{tab:gaps}
\begin{tabular}{lccccccccc}
\hline
 \rule{0pt}{0.30cm}  & Ref. & cubic  &  $P4/nmm$  &  $P4/ncc$ &  Pcnb  &  P2$_1$/n &  $P\bar{1}$  & P2$_1$/c   \\
 \hline
B1-WC  &                                             & 1.5  &  2.12 &  2.15    & 2.65     &  2.85  &  2.98   &  3.28           \\    
GW      & \onlinecite{Johansson2013}            &        &       &          &          &  2.90   &  3.00       &   3.30             \\
PBE0   & \onlinecite {WangFenggong2011}   & 2.25 &  2.28 &          &  3.35     &  3.67  &  3.67    &              \\
HSE06 & \onlinecite{WangFenggong2011}    & 1.67 &  1.71 &          &  2.57     &  2.80  &  2.94    &                 \\
B3LYB  & \onlinecite{WangFenggong2011}   & 1.89 &  1.85 &          &   2.89    &  3.13  &  3.17    &                   \\
Exp 1   & \onlinecite{GonzalezBorrero2010}  &        &           &         &          &  2.75      &          &      &    \\
Exp 2   & \onlinecite {Zheng2011}                 &         &       &          &          & 3.21    &  3.25    &                    \\
Exp 3   & \onlinecite {migas2010}                 &         &  1.75     &          &   2.35    & 2.60    &      &                    \\
\hline
\end{tabular}
\end{table}

\section{Origin of the WO$_3$ phases}

The results presented so far give us confidence that the B1-WC functional reproduces well the experimental measurements so that we can now focus on the structural instabilities of the hypothetical $Pm\bar{3}m$ cubic parent phase and explain how their condensation give rise to the various known phases of WO$_3$.
This also allows us to identify novel ferroelectric metastable phases. In each case, we analyze the crystallographic structure through a decomposition of the distortions with respect to the cubic parent phase in terms of symmetry-adapted modes.

\subsection{Unstable modes of the cubic reference}

\begin{figure}[h]
 \centering
 \includegraphics[width=8.5 cm,keepaspectratio=true]{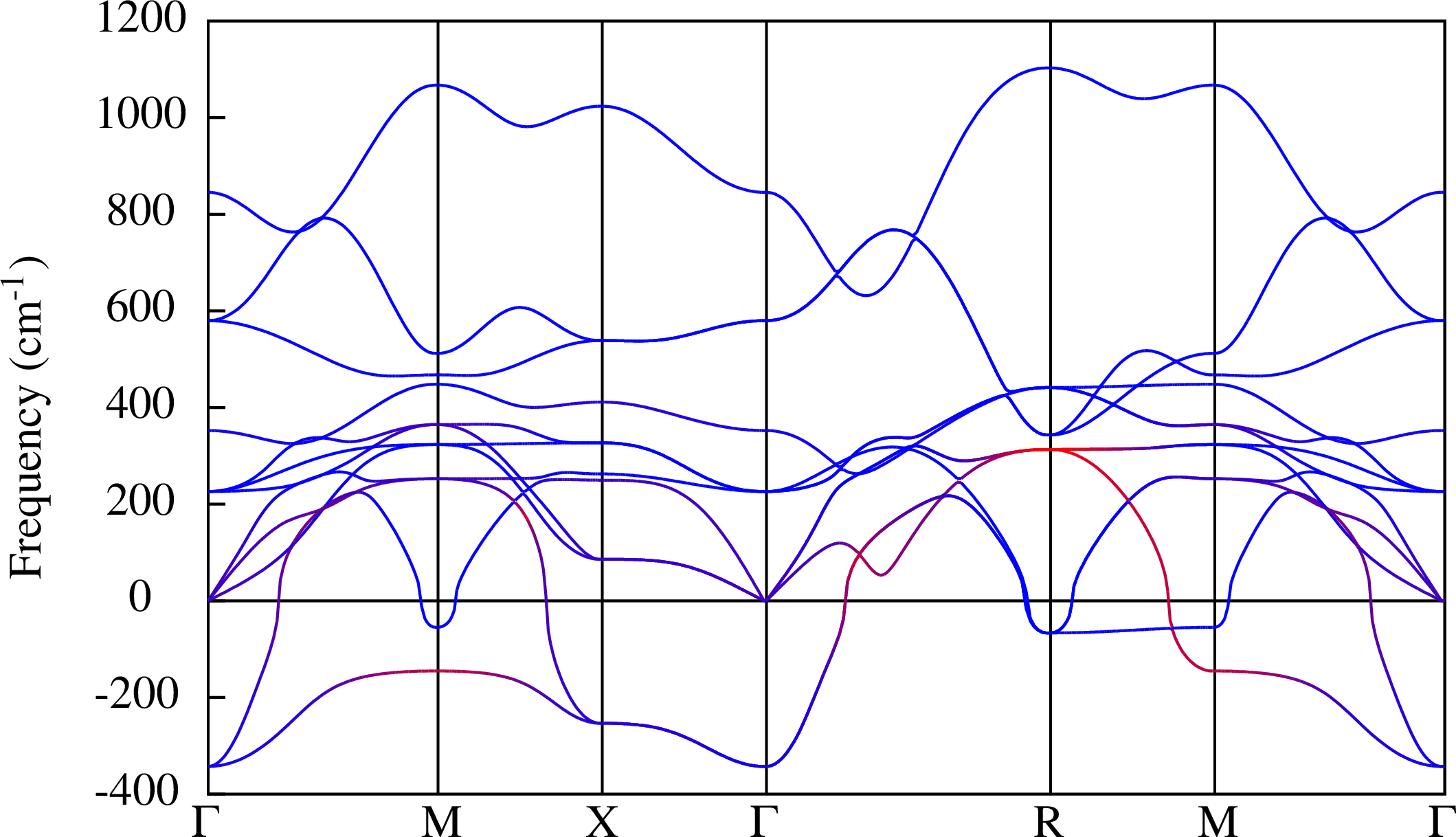}
 \caption{(Color online) Phonon dispersion curves of cubic WO$_3$ (negative frequencies refer to imaginary frequencies, \textit{i.e.} to unstable modes). The coordinates of the high symmetry points are as follows: $\Gamma$ (0,0,0), X $(\frac{1}{2},0,0)$, M $(\frac{1}{2},\frac{1}{2},0)$ and R $(\frac{1}{2},\frac{1}{2},\frac{1}{2})$.
 Thanks to the \textit{band2eps} postprocessing script of ABINIT,\cite{abinit2009} the color of the bands is assigned to each point through the contribution of each atom type to the corresponding eigenvector: red for the tungsten atom and blue for the oxygens.}
 \label{fig:phonon_dispersions}
 \end{figure}

Fig.~\ref{fig:phonon_dispersions} shows the calculated phonon dispersion curves of hypothetical cubic WO$_3$.
Two branches of instabilities (imaginary frequencies plotted as negative numbers in Fig.\ref{fig:phonon_dispersions}) coexist in the Brillouin zone.

The first unstable branch has its largest imaginary value at $\Gamma$. The $\Gamma$ unstable mode has the irreducible representation (irrep)
$\Gamma_4^-$ and corresponds to a polar mode.  It suggests that the cubic phase of WO$_3$ is mostly unstable via this polar instability and might be
ferroelectric, which we will see later is not exactly the case. The polar instability at $\Gamma$ propagates toward the X and M points with weak dispersion
while it strongly  disperses towards the R point.  Aside from $\Gamma$, the modes of this branch are anti-polar.  The dispersion of this unstable branch is
very similar to the one reported in BaTiO$_3$ and corresponds to a ferroelectric instability requiring a chain-like correlation of displacements in real space.\cite{ghosez1999} 

The second branch of unstable modes appears between M and R points with smaller amplitudes and a nearly absent dispersion between these two points.
The label of the M and R point unstable phonon modes are M$_3^{+}$ and R$_4^+$   and they correspond to rotations of the oxygen octahedra.
The dispersion of this branch is comparable to what is observed for similar modes in SrTiO$_3$ or PbTiO$_3$ and linked to a planar character of
the correlations of the atomic displacements in real space. \cite{ghosez1999}

\begin{figure}[htbp!]
 \centering
 \includegraphics[width=8.5cm,keepaspectratio=true]{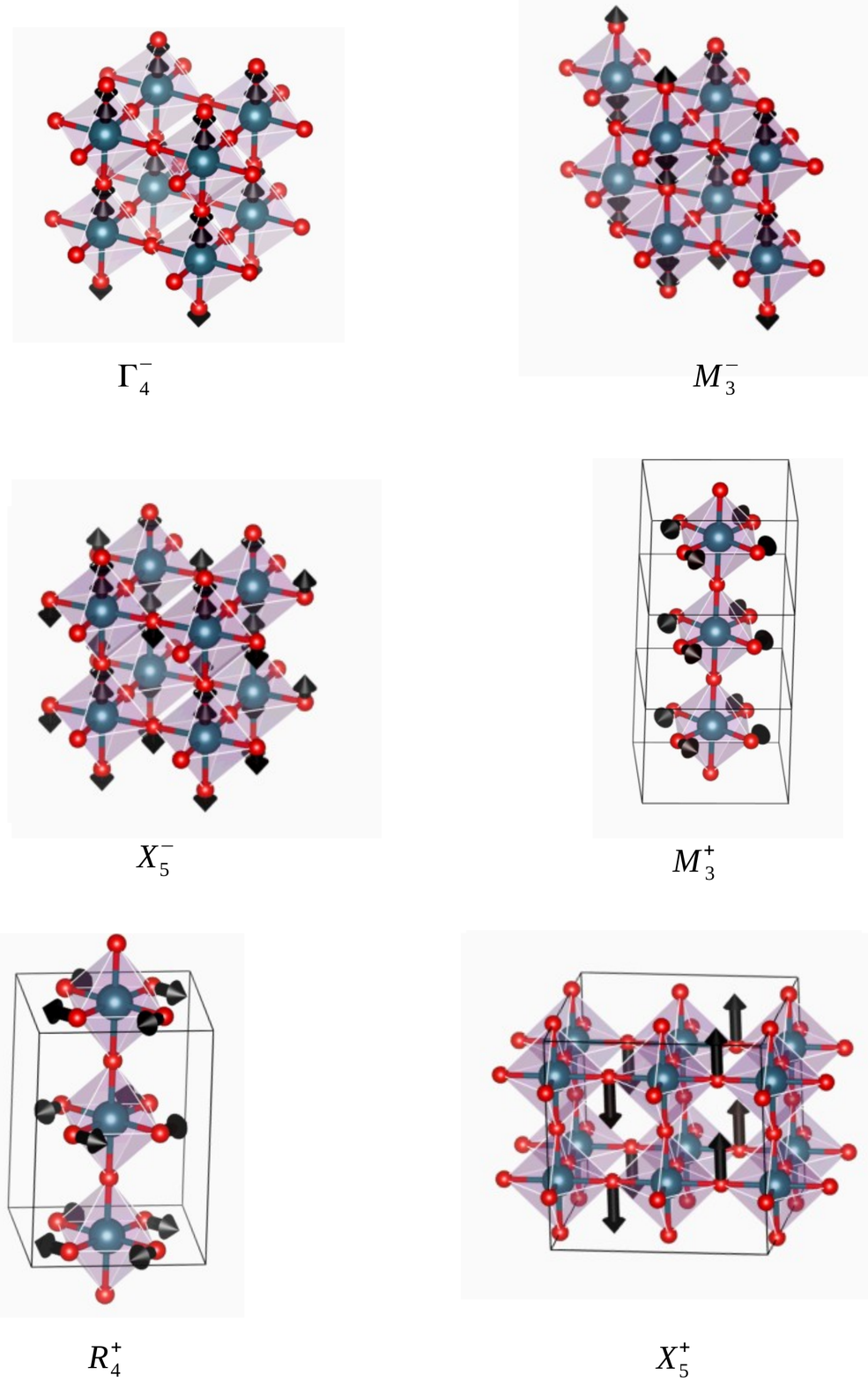}
 \caption{(Color online) Schematic view of most important modes contributing to the distortions of WO$_3$. Small red spheres represent the oxygens and large blue spheres represent the tungsten atoms. All the modes are unstable but the X$_5^+$ mode, which is discussed in section \ref{sectionSymmAdapt}.}
 \label{fig:unstable-modes-VESTA}
\end{figure}

Fig.~\ref{fig:unstable-modes-VESTA} represents a schematic view of the eigenvectors related to the main instabilities of cubic WO$_3$.
The polar mode at $\Gamma$ ($\Gamma_4^-$, 373i cm$^{-1}$) shows motion of W against the O atoms, which is the source of a large electrical polarization.
The anti-polar modes at the X (X$_5^-$, 256$i$ cm$^{-1}$) and M points (M$_3^-$, 147$i$ cm$^{-1}$) are associated to opposite displacements from unit cell
to unit cell along the [100] and [110] directions respectively.
\footnote{In the cubic cell, the [100], [010] and [001] directions are degenerate. The same applies for the [110], [101] and [011] directions.}
The M$_3^{+}$ (62i cm$^{-1}$)  and R$_4^+$ (69i cm$^{-1}$) unstable modes correspond rotations of the oxygen octahedra about the central W atom with consecutive  octahedra along the rotation direction moving respectively in the same or opposite directions.
Using the Glazer notation,\cite{glazer1972} the M$_3^{+}$ mode corresponds to $a^0a^0a^+$ and the R$_4^+$ mode to $a^0a^0a^-$.

\subsection{Condensation and coupling of modes}

Starting from the previous unstable modes, we now investigate how their individual and combined condensations in the hypothetical cubic
structure give rise to various phases. We then compare their energies and analyze the amplitudes of distortions.

\subsubsection{Condensation of modes of the unstable polar branch}

\begin{figure}[h]
 \centering
 \includegraphics[width=10 cm ,keepaspectratio=true]{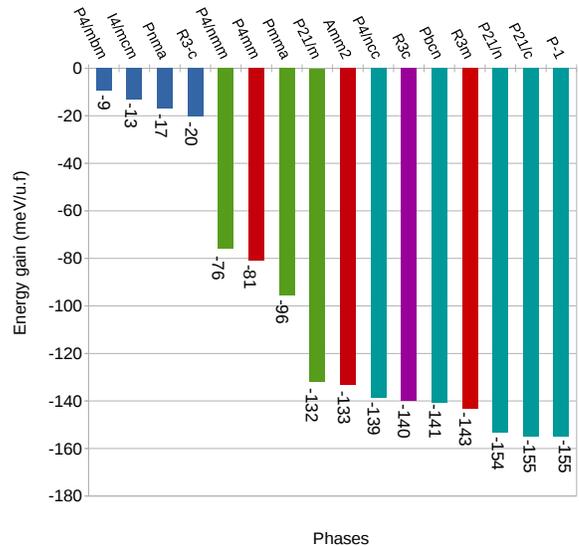}
 \caption{(Color online) Calculated energy gain (in meV/f.u.) with respect to the cubic phase of different phases of WO$_3$. Red columns are the FE phases arising for the condensation of the polar unstable mode, blue columns are the phases arising from the condensation of the oxygens octahedral rotation unstable modes, green columns are the phases arising from the condensation of anti-polar modes, magenta column represents a phase combining FE and anti-polar modes and cyan columns represent phases combining  oxygen octahedral rotations and anti-polar modes. For clarity, the exact value of the energy gain is written in each case.}
 \label{fig:energy-gain}
\end{figure}

We first consider the condensation of unstable $\Gamma_4^-$, X$_5^-$ and M$_3^-$  modes.
Fig.~\ref{fig:energy-gain} shows the energy gain of the corresponding relaxed phase with respect to the cubic phase.
We tested several condensation schemes:
(i) condensation of the polar $\Gamma_4^-$mode along one ($P4mm$), two ($Amm2$) and three ($R3m$) directions ;
(ii) condensation of the X$_5^-$ along one ($Pmma$) and two ($P2_1/m$) directions;
(iii) condensation of the M$_3^-$  mode along one direction ($P4/nmm$).

We observe that the energy gain of the polar instabilities is large and that the $\Gamma_4^-$ polar mode drives a larger gain of energy (red columns in Fig.~\ref{fig:energy-gain}) than the anti-polar X$_5^-$ and M$_3^-$ modes (green columns in Fig.~\ref{fig:energy-gain}).
The space group related to the condensation of the M$_3^-$ mode corresponds to the high temperature phase observed experimentally  ($P4/nmm$).

Condensation of the  $\Gamma_4^-$  mode along two and three directions produce energy gains larger than its condensation in a single direction so that
$\Gamma_4^-$ mode alone will drive the system polar along the [111] direction with an energy difference between the $Amm2$ and $R3m$ phases of 6 meV.
We calculated the polarization amplitude in the three $P4mm$, $Amm2$ and $R3m$ phases using the Berry phase technique and obtain 54, 69 and 69  $\mu$C.cm$^{-2}$.
These polarization values are comparable to those observed in robust ferroelectrics such as PbTiO$_3$.
They can be explained by the opposite motions of W and O atoms, associated with strongly anomalous Born effective charges (11.73 e for W and -8.78/-1.62 e for O$_\parallel$/O$_\perp$ in good agreement with previous calculations in Ref.~\onlinecite{DetrauxF1997}).

\subsubsection{Condensation of modes of the oxygen rotation unstable branch}

Fig.~\ref{fig:energy-gain} (blue columns) shows the energy gain given by the condensation of the M$_3^{+}$ and R$_4^+$  modes along one direction ($I4/mbm$ and $I4/mcm$), the R$_4^+$ mode in three directions ($R\bar{3}c$) and the orthorhombic $Pnma$ phase where the R$_4^+$ mode is condensed in two directions and the  M$_3^{+}$ mode in one direction ($a^-a^-c^+$ ).
These distortions lower the energy much less than the polar and antipolar motions.
This observation is in line with the modest amplitude of the related phonon instabilities : the energy curvatures at the origin are less negative for the octahedral rotations than for the polar motion.
Nevertheless the amplitude of octahedral rotations are 10.7 and 11.7 degrees in $I4/mbm$ and $I4/mcm$, respectively.
Such large distortions associated to a weak instability highlight relatively small anharmonicities, which might be explained by the absence of $A$ cation with respect to regular $A$BO$_3$ perovskites.\cite{Benedek2013}

\subsubsection{Combinations of modes}

Beside the $P4/nmm$ phase, none of the previous single irrep mode condensations correspond to observed phases.
Thus, we now explore the condensation of combined octahedral rotations and polar/anti-polar modes.
We depict in Fig.~\ref{fig:energy-gain} the energy gain given by the joint condensation of polar and oxygen rotation modes along three directions ($R3c$, in purple color) and of anti-polar and oxygen rotation modes ($P4/ncc$, $Pbcn$, $P2_1/n$, $P2_1/c$ and $P\bar{1}$, in cyan color).

Combining the polar distortion of the low energy $R3m$ phase with additional oxygen rotation modes does not further reduce the energy.
Instead, it yields a $R3c$ phase slightly higher in energy but with a slightly amplified polarization of 71 $\mu$C.cm$^{-2}$ and a slightly reduced oxygen rotation (with respect to the $R\bar{3}c$).
This emphasizes an unusual competition between these two types of distortions in WO$_3$ with respect to regular perovskite compounds, where in WO$_3$ the $R3c$ phase forms a local minimum between the $R\bar{3}c$ and $R3m$ phases.

At the opposite, the mixing of the anti-polar mode M$_3^-$ with oxygen rotation modes can drive larger energy gains so that the ferroelectric $R3m$ phase is not the ground state.
This is in agreement with experimental observations where the observed phases at all temperatures contain anti-polar motions.
Amongst investigated phases, the $P2_1/c$ phase appears as the most stable but only marginally, as we observe that the $P2_1/n$, $P2_1/c$ and $P\bar{1}$  phases are all extremely close in energy (energy gains of 153, 155 and 155 meV respectively, see Fig.~\ref{fig:energy-gain}). 
Consequently, within the precision of our calculations, we cannot unambiguously assess which one is the ground state. Nevertheless, as discussed in Section III-A and further exemplified in the next Section, the $P2_1/c$ phase is in excellent agreement with the experimental  $Pc$ ground state, except for a tiny polar distortion. 
Our calculations highlight that, in fact, the $P2_1/n$ and $P\bar{1}$ phases observed at higher temperatures are also extremely close in energy.

We further notice that the ferroelectric $R3m$ phase, although never observed experimentally, is also relatively close in energy to the ground state (about 11 meV/f.u.).
Following K. M. Rabe,~\cite{rabe2013} the non-polar (or eventually weakly polar in the experimental $Pc$ phase) ground-state of WO$_3$ combined with an alternative low-energy ferroelectric phase obtained by polar distortions of the same high-symmetry reference structure makes it a potential antiferroelectric compound.
Indeed, applying an electric field, it might be possible to open a typical double hysteresis loop from a field-induced first-order transition from the $P2_1/c$ ground state  to the $R3m$  polar phase. Estimating the critical electric field required to stabilize the $R3m$ phase from $\mathcal{E}_c \sim \Delta E / \Omega_0 P_s$,~\cite{PhysRevB.90.140103}  where $\Delta E$ is the energy difference between the two phases (11.43 meV/f.u.), P$_s$ the spontaneous polarization  of the polar phase (69$\mu$C.cm$^{-2}$)
and $\Omega_0$ its unit cell volume (55\AA), we get the relatively modest value $\mathcal{E}_c\sim$ 480 kV/cm.
For the polar phase $R3c$ we need to apply a greater electric field $\mathcal{E}_c\sim$ 638 kV/cm to stabilize this phase.
This allows us to estimate that the critical field has smillar value
with respect to other antiferroelectric material, $\mathcal{E}_c\sim$ 470kV/cm for Zr$O_2$ \cite{PhysRevB.90.140103} and $\mathcal{E}_c\sim$ 239kV/cm for PbZr$O_3$. \footnote{Value calculated from the energies and polarization reported for the $R3m$ and $Pbam$ phase by S. Amisi in his \href{http://hdl.handle.net/2268/158512}{PhD thesis}}
Although this might not be easy to check experimentally on real samples that are typically oxygen deficient and highly conductive, the calculations reveal that stoichiometric WO$_3$ exhibits all the features of an antiferroelectric compound. 

\subsubsection{Symmetry adapted mode analysis of the distorted phases}\label{sectionSymmAdapt}

\begin{figure}[h]
 \centering
 \includegraphics[width=8.5cm ,keepaspectratio=true]{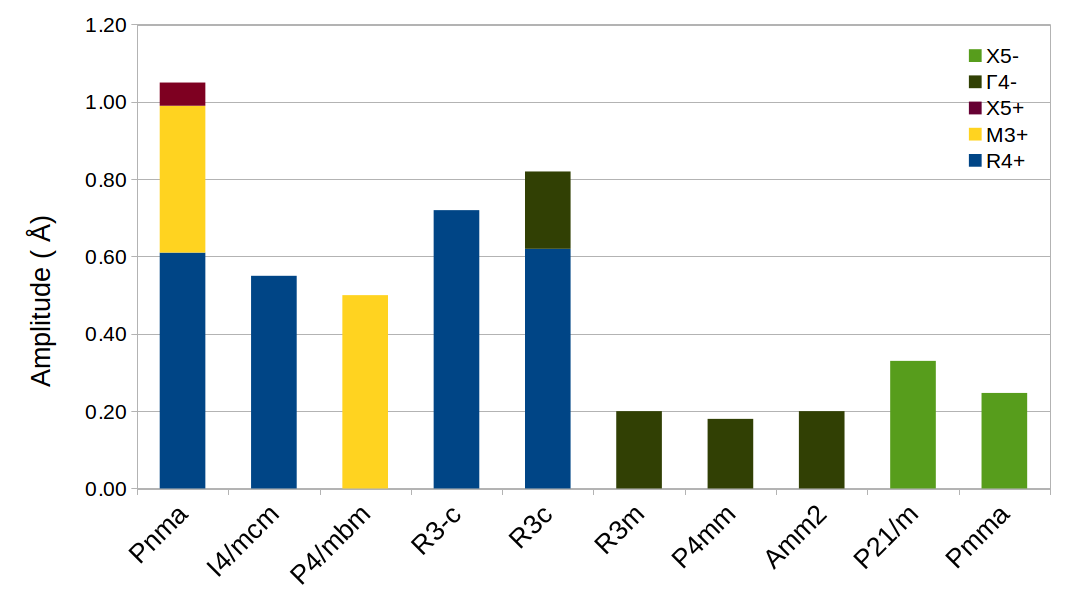}
 \caption{(Color online) Symmetry adapted mode decomposition of distorted phases of WO$_3$ explored in our study but not observed experimentally.}
 \label{fig:amplimode2}
 \end{figure}

\begin{figure}[h]
 \centering
 \includegraphics[width=8.5cm ,keepaspectratio=true]{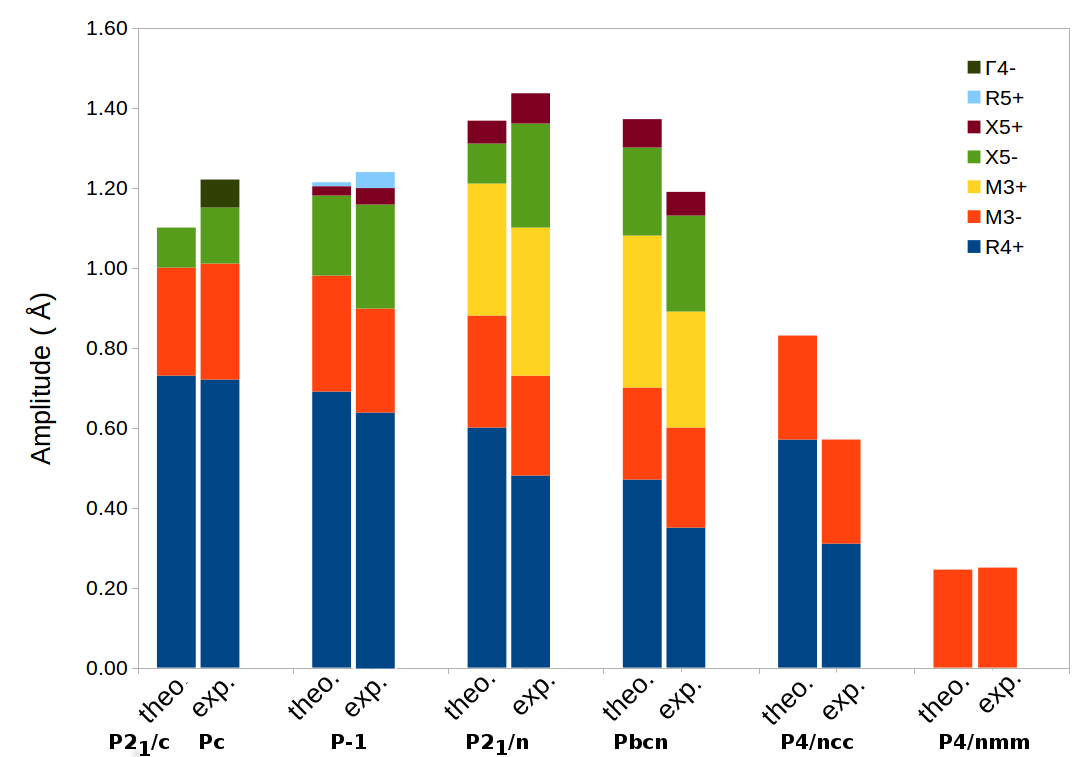}
 \caption{(Color online) Symmetry adapted mode decomposition of distorted phases WO$_3$; Comparison between experiments and our calculations with the B1-WC functional are shown.}
  \label{fig:amplimode1}
\end{figure}

To quantify the distortions that appear in the various phases we project the structural distortions onto symmetry adapted modes of the cubic phase using AMPLIMODE software.\cite{amplimode}
 The results in Fig.~\ref{fig:amplimode2} show the amplitudes of the modes in the fully relaxed phases from the calculations but non-observed experimentally.
In Fig.~\ref{fig:amplimode1} we show the amplitudes of modes in both the fully relaxed and observed phases, which can be compared.

In the following we discuss the competition/cooperation character of the mode distortions.
In perovskite oxydes, it is established that the oxygen rotations are in competition with the ferroelectric displacements but less attention has been given to the combinations of other types of mode.
Often, this cooperation or competition comes from the biquadratic energy term in the free energy expansion with respect to two order parameters.
In WO$_3$, we observe that the combination between the  $\Gamma_4^-$ mode and the R$_4^+$ mode along the [111] direction in the $R3c$ phase has the tendency to reduce the amplitude of the oxygen rotations with respect to the $R\bar{3}c$ phase (the  R$_4^+$ mode is 13\% smaller in the $R3c$ phase than in the $R\bar{3}c$ phase, see  Fig.~\ref{fig:amplimode2}) while the polar mode is unaffected.
As discussed in the previous section, the combination of the  $\Gamma_4^-$ and R$_4^+$ modes forms a local minimum ($R3c$ phase) of higher energy than the $R3m$ phase.
This means that the polar distortions are in competition with the oxygen rotations as reported for perovskite oxydes, with the difference that the polar mode amplitude is unaffected and that the $R3c$ phase is locally stable (the system does not relax into the lowest energy $R3m$ phase).
The strain can also play an important role,\footnote{The volume of the cubic,  $R\bar{3}c$, $R3m$ and $R3c$ phases are 54.01\AA$^3$, 50.89\AA$^3$, 55.35\AA$^3$, 52.71\AA$^3$} but when performing the same calculations at fixed cell parameters (fixed to the cubic one), we find that the $R3c$ phase still forms a local minimum of higher energy than the $R3m$ phase.
This unusual energy landscape can come from the marginal gain of energy of the oxygen rotations while large amplitude of rotations are present.

On the other hand, the association of the oxgen rotations with the antipolar M$_3^-$ mode is cooperative.
When we compare the amplitude of the R$_4^+$ and M$_3^-$ modes of the $P4/nmm$, $I4/mcm$ and $P4/ncc$ (Fig.~\ref{fig:amplimode2} and Fig.~\ref{fig:amplimode1}) we find that when both the R$_4^+$ and M$_3^-$ modes are present together in the $P4/ncc$ phase, their amplitude is slightly higher (4\% larger)  than when condensed alone ($P4/nmm$ and $I4/mcm$ phases).
Their combination, however, drives a sizeable gain of energy: the $P4/ncc$ phase is 63 meV and 126 meV lower in energy than the $P4/nmm$ and the $I4/mcm$ phases, respectively.
This means that the combination of the oxygen rotations with the antipolar M$_3^-$ mode is much more cooperative than the combination with the polar mode $\Gamma_4^-$.

The $Pbcn$ phase can be understood as a distorted $P4/ncc$ phase with additional  M$_3^+$ oxygen rotations along [010].
The resulting tilt pattern is a$^0$b$^+$c$^-$ with a small energy gain of 2 meV with respect to the $P4/ncc$ phase and a reduction of mode amplitudes M$_3^-$, R$_4^+$ and M$_3^+$ (16\%, 7\% and 22\% reduction of the M$_3^-$, R$_4^+$ and M$_3^+$ modes with respect to the phases where they are condensed alone, \textit{i.e.}, $P4/nmm$, $I4/mcm$ and $P4/mbm$, respectively).
The M$_3^+$ mode competes with the R$_4^+$ and M$_3^-$ modes in the sense their combination reduces their amplitude, but they cooperate to lower the energy of the system.

In the case of the monoclinc $P2_1/n$ , $P2_1/c$ and triclinic $P\bar{1}$ phases, the combination of the M$_3^-$ mode with several oxygen
rotations (a$^-$b$^+$c$^-$ for $P2_1/n$, a$^-$a$^-$c$^-$ for $P2_1/c$ and a$^-$b$^-$c$^-$ for $P\bar{1}$) lowers the energy of the crystal and with an increase of the
mode amplitude with respect to the phases where these modes are condensed independently.
For example, the antipolar M$_3^-$ mode has his amplitude increased by 11\%, 10\% and 19\% in the  $P2_1/n$ , $P2_1/c$ and $P\bar{1}$ phases respectively.
This means that the dominant R$_4^+$ oxygen rotations cooperate with the antipolar M$_3^-$ mode to promote the ground state of WO$_3$.

We note that in the $P4/nmm$, $P4mm$, $Amm2$, $R3m$,  $I4/mcm$, $I4/mbm$, $R\bar{3}c$, $P4/ncc$ and $R3c$ phases the mode decomposition shows only the primary modes we have condensed.
This is different with the $Pnma$ phase in which a secondary mode X$_5^+$ appears with a small amplitude in the mode projections while we have condensed only the primary R$_4^+$ and M$_3^+$ modes (see Fig.~\ref{fig:amplimode2}).
This additional mode appears by improper coupling between the R$_4^+$ and M$_3^+$ modes such that the symmetry of the $Pnma$ structure allows the X$_5^+$ mode to develop even though the X$_5^+$ mode is not unstable by itself.\cite{young2015}
Similarly, we observe the apparition of several additional secondary modes in the $Pbcn$, $P2_1/n$ , $P2_1/c$ and  $P\bar{1}$ phases, which we discuss in the next section.

\subsubsection{Energy invariants and improper couplings}

The different phases arise from the condensation of one or several unstable modes (primary modes) of the cubic parent structure but, in some cases, also include the further condensation of stable modes (secondary modes) with significant amplitudes.
Often, the appearance of  such secondary modes of large amplitude can be traced back in their linear coupling with the  primary modes.\cite{bousquet2008,Benedek2015}
This can be rationalized from the existence  in the energy expansion of the $Pm\bar{3}m$ phase of invariant terms of the form $\gamma Q_s\prod_{i=1}^{N}Q_p^i$ where $Q_s$  is the amplitude
of the secondary mode $s$ and $Q^i_p$ the amplitude of the primary mode $i$. Depending of the number of primary mode involved, these coupling terms can be bilinear (pseudo-proper), tri-linear (improper and hybrid-improper),
quadri-linear, etc.

To clarify the origin of secondary modes in several low symmetry phases of WO$_3$, we restrict ourself here to the search of such multi-linear invariant terms in the energy expansion  around its cubic phase by using the INVARIANTS
software.\cite{hatch2003}
In the last column of Table~\ref{tab:mode_decompo_invariants} we report these linear invariants up to the fourth order obtained for the $Pnma$, $Pbcn$, $P2_1/n$ and $P2_1/c$ phases.
The letters represent the mode amplitudes ($Q^i$) in the directions specified in the third column where the bold letters refer to the primary modes in the structure ($Q^i_p$) and the normal letters to secondary modes ($Q_s^i$).

In the $Pnma$ phase, we find that the  X$_5^+$ mode appears through a trilinear coupling with the oxygen rotations modes R$_4^+$ and M$_3^+$ (hybrid-improper coupling, \textbf{ab}c invariant in Table~\ref{tab:mode_decompo_invariants}).
This additional hybrid-improper  X$_5^+$ mode is also found in $Pnma$ of $A$BO$_3$ perovskites~\cite{benedek2012} where the eigenvector corresponds to anti-polar motions of the A cation.
In WO$_3$ the A-cation is absent and X$_5^+$ corresponds to similar anti-polar motions but of oxygen instead of the A-site (see Fig.~\ref{fig:unstable-modes-VESTA}).

In $Pbcn$ the primary M$_3^-$mode condenses along z, the R$_4^+$ mode along z and the M$_3^+$ mode along y and four additional secondary modes:  X$_5^-$ and  X$_5^+$ with a large amplitude and M$_5^+$ and M$_4^+$ with a small amplitude as well as an additional component of the M$_3^-$ mode about the x direction.
If we restrict ourself to the strongest  X$_5^-$ and  X$_5^+$ modes we find that both are coupled with the R$_4^+$ and M$_3^+$ modes through a trilinear coupling (hybrid-improper, \textbf{ab}e and \textbf{ab}f invariants in Table~\ref{tab:mode_decompo_invariants}) but also through a quadrilinear coupling with the R$_4^+$ mode and the two components of the M$_3^-$ mode (\textbf{ad}ce and \textbf{ad}cf invariants in Table~\ref{tab:mode_decompo_invariants}).
We can thus explain the appearance of the X$_5^+$ and X$_5^-$ modes through a trilinear coupling with the oxygen octahedral rotations and
the appearance of the second x component of the anti-polar M$_3^-$mode through a coupling with the secondary  X$_5^+$ and  X$_5^-$ modes and the primary  R$_4^+$ mode.
The final structure can thus be seen as anti-polar through the M$_3^-$ mode along z with a canting of its direction toward the x axis and through the X$_5^-$ mode along the y direction, the whole distortions being associated with the $a^-b^+c^-$ pattern of oxygen rotation distortions.

The transition from $Pbcn$ to $P2_1/n$ can be seen as being induced by the condensation of the R$_4^+$ mode along the remaining direction for the oxygen rotation octahedral distortions to $a^-b^+c^-$.
This means that we find the same mode coupling as in the $Pbcn$ phase plus some extra ones due to the additional mode condensation.
Because we do not induce a new irrep, the couplings are the same (i.e. trilinear and quadrilinear between the primary  R$_4^+$ , M$_3^+$ and M$_3^-$ modes and the  secondary X$_5^-$ and  X$_5^+$ modes) but in different directions from the $Pbcn$ phase: we observe the  X$_5^-$ and  X$_5^+$ modes in two directions instead of one.
Other modes also appear in the symmetry adapted mode analysis but with smaller amplitudes (M$_4^+$, M$_5^-$, M$_5^+$ and M$_2^+$ ), which we do not include in the invariant analysis.

 $P\bar{1}$ and $P2_1/c$ are very similar in the sense that for both structures we can envisage the condensation of R$_4^+$ modes in three directions and the M$_3^-$mode in one direction.
 The difference is that in the $P2_1/c$ phase the R$_4^+$ mode is primary with the same amplitude in two directions and a different amplitude in the third direction ($a^-a^-c^-$where the M$_3^-$mode is primary  in the z direction) while in the  $P\bar{1}$ phase the condensation of the R$_4^+$ mode has different amplitudes in three directions ($a^-b^-c^-$).
In $P2_1/c$ the presence of the  X$_5^-$ and M$_5^-$ secondary modes can be explained by trilinear coupling with the R$_4^+$ and M$_3^-$ primary modes (\textbf{ac}d+\textbf{bc}d  and \textbf{ab}e invariants in Table~\ref{tab:mode_decompo_invariants}) in a similar way as in the $P2_1/n$ phase.

This analysis shows that the low symmetry phases of WO$_3$ are complex and involve numerous multilinear couplings of modes if one expands the energy about the cubic phase.
We note that, amongst possible couplings, the coupling with the secondary X$_5^-$ mode is most important in all low symmetry phases.

Going further, in order to test whether symmetry arguments can lead to the the polar $Pc$ phase using improper-like couplings, we have tested if there exists a coupling with a polar mode at the $\Gamma$  point.
We did not find any couplings with the R$_4^+$ , M$_3^+$ or M$_3^-$ modes.
We thus conclude that it is not possible to generate polarity in WO$_3$ in the limit to these primary  modes, which are the ones appearing in other experimental phases.
\begin{table*}
\centering
 \begin{tabular}{cclccccc}
   \hline
  \hline
 \multirow{2}{*}{Space group} &   \multirow{2}{*}{Irrep.} & \multirow{2}{*}{Direction} & \multirow{2}{*}{Subgroup} & \multicolumn{2}{c}{Amplitudes (\AA) \rule{0pt}{0.35cm}} & \multirow{2}{*}{Linear Invariants} \\
                      &             &                 &                  & Calc.  & Exp. & \\
  \hline
\multirow{3}{*}{$Pnma$}  \rule{0pt}{0.35cm}
  & \textbf{R$_4^+$}   & (\textbf{a},\textbf{a},0)                         & $I4/mma$       & 0.61    & ---  &   \\
  & \textbf{M$_3^+$}  & (0,0,\textbf{c})                         & $P4/mbm$     & 0.38  & ---   &  \textbf{ab}c \\
  & X$_5^+$   & (0,a,0,0,0,0) & Cmcm & 0.06 & --- &\\
  \hline
\multirow{5}{*}{$Pbcn$}  \rule{0pt}{0.35cm}
  & \textbf{R$_4^+$}   & (0,0,\textbf{a})           & $I4/mcm$     & 0.47 & 0.35 &   \\
  & \textbf{M$_3^+$}  & (0,\textbf{b},0)           & $P4/mbm$    & 0.39 & 0.30 & \textbf{ab}e, \textbf{ad}ce  \\
  & \textbf{M$_3^-$}   & (c,0,\textbf{d})           & $Ibam$      & 0.23 & 0.25 & \textbf{ab}f, \textbf{ad}cf \\
  & X$_5^-$   & (0,0,e,-e,0,0) & $Pmma$        & 0.22 &  0.25  &  \\
  & X$_5^+$  & (0,0,0,0,f,f)    & $Pmma$         & 0.07 & 0.06  & \\

  \hline
  \multirow{5}{*}{$P2_1/n$}  \rule{0pt}{0.35cm}
  & \textbf{R$_4^+$}   & (0,\textbf{a},\textbf{b}) \rule{0pt}{0.35cm}            & $C2/m$        & 0.60 & 0.48 &\textbf{ae}g+\textbf{bd}f \rule{0pt}{0.35cm} \\
  & \textbf{M$_3^+$}  & (\textbf{c},0,0)             & $P4/mbm$    & 0.34 & 0.37 & \textbf{ac}h+\textbf{ac}i-\textbf{bc}h+\textbf{bc}i\\
  & \textbf{M$_3^-$}   & (0,\textbf{d},\textbf{e})             & $Ibam$         & 0.27 & 0.25 & \textbf{acd}g$-$\textbf{bce}f  \\
  & X$_5^-$    & (f,-f,0,0,g,g)   & $Pmmn$       & 0.09 & 0.26 & \textbf{ade}h+\textbf{ade}i+\textbf{bde}h$-$\textbf{bde}i\\
  & X$_5^+$   & (0,0,h,i,0,0)    & $P2_1/m$     & 0.06 & 0.08 & \\
  \hline
  \multirow{4}{*}{$P2_1/c$}  \rule{0pt}{0.35cm}
  & \textbf{$\Gamma_4^+$} & (\textbf{a},-\textbf{a},-\textbf{b})         & $Cm$            &  0 & 0.067 &   \\
  & \textbf{R$_4^+$} & (-\textbf{b},\textbf{a},-\textbf{a})         & $C2/c$            & 0.73 & 0.72 & \textbf{ac}d+\textbf{bc}d \\
  & \textbf{M$_3^-$}  & (\textbf{c},0,0)           &  $P4/nmm$      & 0.27 & 0.3 & \\
  & X$_5^-$   & (0,0,0,-d,0,0) & $Cmcm$         & 0.10 & 0.14 &  \\
  \hline
  \hline
 \end{tabular}
\caption{Symmetry adapted modes decomposition and linear couplings of modes of the $Pnma$, $Pbcn$, $P2_1/n$ , $P2_1/c$ and $P\bar{1}$  phases. From the left to right columns, we show the mode label (Irrep.) of the symmetry adapted mode, the direction of the mode condensation, the corresponding subgroup, the amplitude of the distortion in the calculated and in the experimental cases (the modes with an amplitude lower than 0.005 \AA\ are not shown) and the linear coupling invariants of the most relevant modes where the letters correspond to the one given in the direction column (we highlight in bold the primary  modes).}
\label{tab:mode_decompo_invariants}
\end{table*}

\section{Conclusion}

In this study, we have performed a first-principles study of WO$_3$ using the B1-WC hybrid exchange-correlation functional and highlighted that, in comparison to various other functionals, it yields the best overall agreement with experiments regarding electronic and structural properties.

Starting from the inspection of the phonon dispersion curves of an hypothetical cubic structure taken as reference, we have identified two main branches of instabilities and characterized various phases arising from the condensation of one or more unstable modes.  Although the dominant phonon instability is associated to a zone-center polar mode, we found a non-polar $P2_1/c$ ground state arising from the combination of cooperative anti-polar distortions and oxygen octahedra rotations. This phase is very similar to the experimentally reported polar $Pc$ ground state, except for the absence of a tiny polar distortion. Our calculations does not show however any tendency of the  $P2_1/c$  phase to evolve to a $Pc$ phase suggesting that WO$_3$ is likely not intrinsically ferroelectric. Instead the ferroelectric character might arise from extrinsic defects such as oxygen vacancies. The $P2_1/c$ phase is anti-polar and defects could easily produce a slightly unbalanced anti-dipole structure, yielding a weak net polarization. In this sense, off-stoichiometric WO$_3$ might be better described as a {\it ferrielectric} compound.\cite{pulvari1960}
The ground state is determined by two antiparallel movements of W off-centerings which exactly compensate each other in the $P2_1/c$ phase.
The displacements are almost identical in the $Pc$ phase, but the two displacements do not fully compensate each other.
We suspect that such weak ferrielectricity can be induced by defects such as oxygen vacancies.

At the level of our calculations, the $P2_1/c$ ground-state is almost degenerated in energy with the $P2_1/n$ and $P\bar{1}$ phases observed at higher temperature. Also, we discovered the existence of a never observed and low-energy ferroelectric $R3m$ phase with a large polarization. Although this might not be of direct interest due to the conductive character of usual off-stoichiometric samples, the proximity with the $P2_1/c$ ground-state of this structurally-unrelated $R3m$ polar phase toward which the system could be switch through a first-order transition under moderate electric fields, makes WO$_3$  a potential antiferroelectric material.
Investigations of the influence of polarons on the WO$_3$ structures are currently underway.

\section{aknowledgements}
We thank Julien Varignon for his helpful guidance with the CRYSTAL code and Emilio Artacho for interesting discussions.
H. H., E. B. and Ph. G. acknowledge the Consortium des Equipements de Calcul Intensif (CECI), funded by the FRS-FNRS (Grant 2.5020.11 and and No. 1175545), and the PRACE project TheDeNoMo for computing facilities and the ARC project AIMED for financial support. E. K. H. S. is grateful for support to EPSRC and the Leverhulme trust. H. H. thanks the AVERROES-ERASMUS Mundus project.

%
\end{document}